\documentclass[a4paper,fleqn,usenatbib,onecolumn]{mnras}

\usepackage[T1]{fontenc}
\usepackage{ae,aecompl}

\usepackage{graphicx}	
\usepackage{amsmath}	
\usepackage{amssymb}	
\usepackage{pdflscape}

\usepackage[usenames,dvipsnames]{color}          

\DeclareFontFamily{OT1}{pzc}{}
\DeclareFontShape{OT1}{pzc}{m}{it}%
             {<-> s * [1.1500] pzcmi7t}{}
\DeclareMathAlphabet{\mathscr}{OT1}{pzc}%
                                 {m}{it}

\newcommand{\half}{\frac{1}{2}}
\newcommand{\threeontwo}{{\textstyle\frac{3}{2}}}

\newcommand{\re}{\mathop{\rm Re}\nolimits}

\newcommand{\sgn}{\mathop{\rm sgn}\nolimits}

\renewcommand{\b}{{\bmath{b}}}
\newcommand{\e}{\hat{\bmath{e}}}

\newcommand{\boldv}{{\bmath{v}}}

\newcommand{\B}{{\bmath{B}}}

\newcommand{\F}{{}_2F_3}

\renewcommand{\P}{{\bmath{P}}}

\renewcommand{\S}{\mbox{\boldmath$\mathsf{S}$}}

\newcommand{\vcross}{{\boldsymbol{\times}}}

\newcommand{\kpar}{k_{\scriptscriptstyle\parallel}}

\newcommand{\eperp}{\e_{\scriptscriptstyle\!\perp}}
\newcommand{\epar}{\e_{\scriptscriptstyle\parallel}}

\newcommand{\ri}{\mathrm{i}}

\newcommand{\ts}{\textstyle}

\newcommand{\kp}{\kappa_0}
\newcommand{\kpz}{\kappa_z}

\newcommand{\BV}{Brunt-V\"ais\"al\"a}

\renewcommand{\leq}{\leqslant}  \renewcommand{\le}{\leqslant}
\renewcommand{\geq}{\geqslant}  \renewcommand{\ge}{\geqslant}

\title[Mode Conversion of Waves from Above]{On Mode Conversion, Reflection and Transmission of Magneto-Acoustic Waves from Above in an Isothermal Stratified Atmosphere}

\author[S. C. Hansen et al.]{
Shelley C.~Hansen,$^{1,\,2}$\thanks{E-mail: shansen@trinity.unimelb.edu.au (SCH)}
Paul S.~Cally,$^1$\thanks{E-mail: paul.cally@monash.edu (PSC)}
Alina-C. Donea$^1$\thanks{E-mail: alina.donea@monash.edu (ACD)}
\author[S. C. Hansen et al.]{
Shelley C.~Hansen\altaffilmark{1}, Paul S.~Cally, and Alina Donea}
\\
$^1$School of Mathematical Sciences and Monash Centre for Astrophysics, Monash University, Clayton, Victoria 3800, Australia\\
$^2$Trinity College, University of Melbourne, Royal Parade, Parkville, Victoria 3052, Australia
}

\date{}
\pubyear{2015}

\begin{document}
\label{firstpage}
\pagerange{\pageref{firstpage}--\pageref{lastpage}}
\maketitle

\begin{abstract}
We use the exact solutions for magnetoacoustic waves in a two dimensional isothermal atmosphere with uniform inclined magnetic field to calculate the wave reflection, transmission, and conversion of slow and fast waves incident from above ($z=\infty$). This is relevant to the question of whether waves excited by flares in the solar atmosphere can penetrate the Alfv\'en/acoustic equipartition layer (which we identify as the canopy) to reach the photosphere with sufficient energy to create sunquakes. It is found that slow waves above the acoustic cutoff frequency efficiently penetrate (transmit) as acoustic (fast) waves if directed at a small attack angle to the magnetic field, with the rest converting to magnetic (slow) waves, in accord with Generalized Ray Theory. This may help explain the compact nature of seismic sources of sunquakes identified using seismic holography. The incident slow waves can also efficiently transmit at low frequency in inclined field due to the reduction in acoustic cutoff frequency (ramp effect). Incident fast (magnetic) ``waves'' from infinity with specified nonzero horizontal wavenumber are necessarily evanescent, but can carry energy to the equipartition level by tunnelling. It is found that this can then efficiently convert to acoustic (fast) energy that can again reach the photosphere as a travelling wave. Overall, there appear to be ample avenues for substantial compressive wave energy to penetrate the canopy and impact the photosphere.
\end{abstract}



{\section{Introduction: Waves in the Solar Atmosphere}

Innumerable papers have explored upward propagation of waves in the solar atmosphere from the photosphere, largely in attempts to explain atmospheric heating \citep[reviewed by][for example]{Asc06aa,ParDe-12aa,Arr15aa}. Other authors have emphasised the role of wave propagation direction relative to the magnetic field vector \citep{Cal06aa,SchCal06aa,Cal07aa, HanCal09aa}, using a variety of approximate mathematical and numerical techniques. In particular, transmission (acoustic-to-acoustic or magnetic-to-magnetic) across the Alfv\'en-acoustic equipartition level $a=c_s$, where $a$ is the Alfv\'en speed and $c_s$ the sound speed, is highly favoured by a small attack angle between the wavevector and magnetic field. Interestingly, there is one valuable and relevant model that admits \emph{exact} solutions: that of an isothermal gravitationally stratified atmosphere with uniform inclined magnetic field in two dimensions (2D) \citep{ZhuDzh84aa,Cal01aa,Cal09aa}, where these ideas can be confirmed and explored. These exact solutions provide precise formulae for coupling coefficients, quantifying mode conversion between fast and slow magneto\-acoustic waves.

However, less attention has been paid to downgoing magneto\-acoustic waves. Nevertheless, these arise in several contexts, and may be relevant to phenomena observed at photospheric levels. For example, upward travelling fast waves almost inevitably reflect from the steep Alfv\'en speed gradient with height in the solar atmosphere, and re-enter the photosphere, where they are observed as high-frequency halos about active regions \citep{KhoCol09aa,RijMorPrz15aa,RijRajPrz15aa}.

Waves can also propagate along coronal loops from one footpoint to the other. Typically, slow waves  are observed close to the footpoints travelling upward only, but fast waves can traverse the entire loop length \citep{NakVer05aa}.

Waves of various types may also be generated in the corona. For example, it was suggested by \citet{AxfMcK92aa} that high frequency waves can be produced in `microflare' reconnections, and full-scale flares clearly produce widespread and powerful oscillations (\emph{e.g.}, \citealt{WarVrsMag04aa,WarVrsMag04ab}, who interpret EIT waves as fast MHD shocks, though see \citealt{WilDeFSte07aa} for a contrary view). Waves from flares are discussed at greater length in Section \ref{flare}.

The purpose of this article is to explore fast/slow mode conversion using the exact \cite{ZhuDzh84aa} isothermal-atmosphere solutions, but with a specific focus on downgoing waves. The exact connection coefficients are calculated and tabulated in Table \ref{tab:a}, and graphics presented summarizing extensive parameter surveys of wavenumber (specifying in particular attack angle) and frequency. These results generalize those of \citet{McDHoo07aa}, who explored downward propagation through an isothermal atmosphere with vertical magnetic field only, using numerical simulation and an approximate local perturbation technique (in small horizontal wavenumber). Since mode conversion depends crucially on attack angle, the vertical field assumption is very limiting.

The price to be paid for exact solutions though is high. Only linear waves are considered, and only in the two-dimensional (2D) vertical plane containing the uniform magnetic field. And the atmosphere is assumed isothermal. Alfv\'en waves are not included: these incompressive disturbances suffer their own mode conversion -- with the fast wave \citep{CalGoo08aa,CalHan11aa,KhoCal11aa,KhoCal12aa,Fel12aa} -- through an intrinsically three-dimensional (3D) process not picked up by the 2D model introduced below. Nevertheless, although very far from a realistic model of the solar atmosphere, the exact solutions provide useful insights into the underlying physics of how magneto-acoustic waves from above penetrate (or reflect from) the $a=c$ equipartition level, and valuable benchmarks for numerical simulation codes. This work is offered here as a resource rather than as a definitive explanation of any specific solar phenomenon.
}

 \section{Flares and Quakes}\label{flare}
 Notwithstanding our qualms about the applicability of our simplistic model to real solar phenomena, we now briefly address, in a speculative spirit, its possible relevance to sunquakes excited by some but not all solar flares.


Solar flares are a sudden release of energy in the coronal plasma triggered by magnetic reconnection \citep{JudKleDon14aa}.  The primary source of the observed energy is most likely to be the solar magnetic field in strong-field, low-$\beta$ active regions 
{\citep{EmsDenShi12aa}}.  Recently, \cite{MilKerDen14aa} estimated that the total non-thermal electron energy in the X-class solar flare, SOL2011-02-15T01:56 was greater than $2 \times10^{31}$ erg.  Only 15\% of the total non-thermal energy  ($\sim3\times 10^{30}$ erg) was detected in the lower solar atmosphere.    Interestingly, the detected seismic energy integrated over all frequencies (2--10 mHz) in sunquakes represents only a few hundredth or even a thousandth part of the total solar flare energy \citep{DonLin05aa}.  This tiny amount must be efficiently directed and focussed towards the photosphere to trigger seismic ripples.
 
Some flares give rise to observable sunquakes, but the majority, regardless of size, do not \citep{Don11aa}.  Although there have been extensive observations of flares with accompanying seismic transients \citep[see for example][]{KosZha98aa,DonBraLin99aa,DonLin05aa,SychKarAlt15aa}, the exact mechanism that couples flares and sunquakes is still unknown \citep{Don11aa}. Possible mechanisms include; ablation of the chromosphere by high-energy electrons and subsequent generation of a chromospheric shock that travels to the photosphere \citep{Kos14aa}; direct high-energy proton penetration to the photosphere \citep{DonLin05aa}; radiation-linked mechanisms such as back-warming \citep{DonBesCal06aa}; and the sudden reconfiguration of the magnetic structure following the flare and its reconnection event, the ``McClymont magnetic jerk'' \citep[see for example][]{HudFisWel08aa}.  However, all of these suggestions prove problematic in some way \cite[see][for details]{MarDonCal08aa}.

    Following an X-class flare on January 15th 2005, one of the largest sunquakes occurred in AR10720.  \cite{MorDonLin07aa} and \cite{MarDonCal08aa} studied this sunquake extensively using helioseismic holography \citep{LinBra00aa}.   In a seismic map of ``egression'' power \citep{LinBra97aa}, the seismic sources at the solar surface (solarquakes) acting at  6 mHz were found to be located mainly in penumbrae.  This is consistent with numerous other observations that report sunquakes orginating from $\delta$-configuration spots that dominate active regions  \citep{MorDonLin07aa,Don11aa}.  
    In the case of the January 15th seismic event, the central compact acoustic source was 10 Mm in size and associated with the magnetic neutral line of the active region.  This suggests that in some way, the $\delta$-sunspot opened a ``window'' for the flare power, thereby allowing a sudden strike of energy on the lower photosphere in a region of almost horizontal magnetic field. The directionality of wave transmission through the magnetic canopy relative to the magnetic field (as described below) raises the question of whether this acts as a filter mechanism for flare waves that may have implications for sunquake excitation. (The term ``Magnetic Canopy'' means different things to different authors. In the context of wave penetration and the associated change in wave nature, it is generally identified with the $\beta\sim1$ layer  \citep{BogCarHan03aa,FinJefCac04ab,McDHoo07aa}, or more properly $a\sim c_s$. These two definitions differ little, since $\beta=2c_s^2/\gamma a^2$, where $\gamma$ is the ratio of specific heats.)
    
{
The role of the fast MHD wave in the flare process itself has been explored by \cite{LonTar12aa}, for whom it is an essential mechanism propagating reconnection away from an initial current sheet. Without this process, they argue, there is insufficient magnetic energy close enough to the current sheet to adequately supply the flare. We presume that the fast wave will continue to propagate, roughly isotropically, away from the flare site, including downward towards the photosphere. 

Being a 2D model, the \cite{LonTar12aa} calculation does not include the Alfv\'en wave, which decouples. One might expect that the Alfv\'en wave should also be generated in a reconnection event. In a low-$\beta$ plasma like the solar corona, the fast and Alfv\'en waves both propagate at the Alfv\'en speed, though the fast wave does so more-or-less isotropically whilst the Alfv\'en wave propagates strictly along the magnetic field only. In either case, with low coronal Alfv\'en speeds of order several hundred kilometres per second \citep{WarMan05aa}, or much more at greater heights \citep{RegPriHoo08aa,FleHud08aa}, the disturbance can propagate rapidly to the lower atmosphere within a few seconds. There is also the possibility that the fast and Alfv\'en waves can couple where the magnetic field geometry and wavevector are favourably arranged \citep{CalHan11aa}. It is even possible for waves to oscillate between the fast and Alfv\'en states mediated by Hall current, though in practice this will only be significant at much higher frequencies (Hz) than are of interest to us here, and very low in the atmosphere where the ionization fraction is small \citep{CalKho15aa}.

The linearity of our model is obviously inappropriate in a flare region. If EIT waves are fast MHD shocks \citep{WarVrsMag04aa,WarVrsMag04ab}, they are certainly not linear. More profoundly, the flare itself very quickly substantially heats and makes dynamic the underlying atmosphere through which any waves travel, so the fixed isothermal atmosphere assumption becomes invalid  \citep{GraHanFle13aa}. Nevertheless, despite these caveats, let us see where the model takes us.
}

\section{Model and Rationale}
We explore a highly simplified two dimensional (2D) infinite model of the downward propagation of flare-induced waves originating at height $z=\infty$ in an isothermal gravitationally stratified magneto-atmosphere. Of course a flare is a very non-steady event taking place high in the corona typically over several minutes  and spread over variable distances horizontally. The event will inject power at a wide range of frequencies and horizontal length scales . These may be considered as separate Fourier modes proportional to $\exp[\ri(k_xx-\omega t)]$, where $x$ is the horizontal position,  $\omega$ is angular frequency and $k_x$ is horizontal wavenumber. 
The waves are also, unrealistically,  assumed linear for simplicity. The third dimension, $y$, is ignored, thereby excluding coupling to the Alfv\'en wave \citep{CalGoo08aa,CalHan11aa,KhoCal11aa,HanCal12aa}.

Flares typically occur in regions of very low plasma-$\beta$ (the ratio of gas to magnetic pressure). Under those circumstances, slow waves are essentially field-guided acoustic waves, with dispersion relation $\omega^2=\omega_c^2\cos^2\theta+c_s^2\kpar^2$, where $c_s$ is the uniform sound speed, $\omega_c=c_s/2H$ is the acoustic cutoff frequency, $H$ is the density scale height, $\theta$ is the magnetic field inclination from the vertical, and $\kpar=k_x\sin\theta+k_z\cos\theta$ is the component of the wave vector $\bmath{k}$ parallel to the magnetic field \citep{SchCal06aa}. They may propagate ($\kpar$ real) at any frequency in excess of the ramp-reduced acoustic cutoff $\omega_c\cos\theta$ and for any $k_x$.

To leading order, the fast waves' low-$\beta$ dispersion relation is simply $\omega^2=a^2k^2$, where $a$ is the Alfv\'en speed and  $k=|\bmath{k}|$, i.e., $k_z^2=\omega^2/a^2-k_x^2$. Since $\omega^2/a^2$ is very small, they propagate with very large wavelength $\lambda=2\upi/k$. In an infinite atmosphere where $\omega/a\to0$ as $z\to\infty$, as assumed here, there are no propagating fast waves originating from infinity. On the other hand, a source at large $z$ with horizontal spatial structure that imposes a $k_x$ (or more properly a distribution of horizontal wave numbers $k_x>\omega/a$), generates waves that are vertically evanescent. These can carry energy over finite distances by tunnelling, which we explore below (as well as via a transient, which we ignore). 

A further simplification is that we do not model the chromosphere/corona transition region (TR) or corona. Although we could feasibly couple together two isothermal layers representing chromosphere and corona separately, we postpone that study for another time. In effect, therefore, we are just modelling the chromosphere, which is reasonably isothermal (around $10^4$ K), with the incident wave thought to be driven by oscillations of the TR caused in turn by the flare in the corona. {Active region chromospheric Alfv\'en speeds typically reach several tens of $\rm km\,s^{-1}$ or more, compared to the sound speed of less than 10 $\rm km\,s^{-1}$, so an active region atmosphere is already low-$\beta$ before reaching the corona. For example, in the sunspot atmosphere constructed by \citet{PrzSheCal15aa} (their Fig.~1), the $a=c_s$ level ranges in height from about $-400$ km at spot centre to around 300 km at horizontal radius 20 Mm, where $a>10c_s$ above about 800 km.  With a chromospheric thickness of around 2 Mm, there is plenty of low-$\beta$ chromosphere above these levels.}

The model therefore is very idealised. Nevertheless, provoked by the well-known identification of the Alfv\'en/acoustic equipartition layer $a\approx c_s$ as the site of mode conversion between fast and slow waves \citep{SchCal06aa}, we simply ask \emph{``Do downward-propagating slow and fast waves penetrate the equipartition layer, and what forms do they then take?''} Although our results do not provide a complete answer to the question about flare waves, they are relevant to it, and provide a context for interpreting observations and simulations.

\section{Exact Solutions}

We assume a gravitationally stratified isothermal atmosphere with sound speed $c_s$ and density $\rho\propto e^{-z/H}$ where $H$ is the uniform scale height. The Alfv\'en speed increases exponentially with height $a\propto e^{z/2H}$. The magnetic field is uniform and inclined at angle $\theta$ in the $x$-$z$ plane. The velocity perturbations are taken to have $\exp[\ri(k_xx-\omega t)]$ dependence on horizontal position $x$ and time $t$. It is convenient to non-dimensionalize the linearized wave equations following \citet{Cal09aa} and \citet{HanCal09aa}, generalizing \citet{Cal01aa}. Specifically, $\nu=\omega H/c_s$ is frequency, $\kappa=k_xH$ is horizontal wavenumber, and $\zeta=\omega H/a\propto e^{-z/2H}$ will be a scaled vertical position variable that ranges from $\zeta=0$ at $z=\infty$ to $\zeta=\infty$ at $z=-\infty$. It is also convenient to introduce $\kappa_0=\sqrt{\nu^2\sec^2\theta-1/4}$ and $\kappa_z =  \sqrt{\nu^2+(n^2-\nu^2)\kappa^2/\nu^2-1/4}$, where $n=\sqrt{\gamma-1}/\gamma\leq\half$ is the dimensionless {\BV} frequency and $\gamma$ is the ratio of specific heats. The acoustic cutoff frequency $\omega_c=c_s/2H$ corresponds to $\nu_c=\half$. Principal square roots are assumed throughout.

In terms of these variables, the governing fourth-order ordinary differential equation for the component of velocity $u$ perpendicular to the magnetic field is set out in \citet{Cal09aa}, Equation (7), and the general solution is
\begin{equation}
\begin{split}
u = & C_1 u_1+C_2u_2+C_3u_3+C_4u_4\\[6pt]
= & C_1\, \zeta^{-2\kappa}\, 
\F\bigl(\ts\half - \kappa - \ri \kpz,\, \half - \kappa + \ri \kpz;
 \, 1-2\kappa,\, \half - \kappa - \ri\kp-\ri\kappa\tan\theta,\, \half - \kappa + \ri\kp-\ri\kappa\tan\theta;
\, -\zeta^2\sec^2\theta\bigr) 
\\[6pt]
&+ C_2\, \zeta^{2\kappa} \, 
\F\bigl(\ts\half + \kappa - \ri \kpz,\, \half + \kappa + \ri \kpz;
\, 1+2\kappa,\, \half + \kappa - \ri\kp-i\kappa\tan\theta,\, \half + \kappa + \ri\kp-\ri\kappa\tan\theta;
\, -\zeta^2\sec^2\theta\bigr) 
\\[6pt]
&+ C_3\, \zeta^{1-2\ri\kp+2\ri\kappa\tan\theta} \, 
\F\bigl(\ts1-\ri\kp-\ri \kpz+\ri\kappa\tan\theta,\,1-\ri\kp+\ri \kpz+\ri\kappa\tan\theta;
\\   & \qquad\qquad\qquad\qquad\qquad\qquad
\,1-2i\kp,\,\threeontwo-\ri\kp-\kappa+\ri\kappa\tan\theta,\,\threeontwo-\ri\kp+\kappa+\ri\kappa\tan\theta;
\,-\zeta^2\sec^2\theta\bigr) 
\\[6pt]
&+ C_4\, \zeta^{1+2\ri\kp+2\ri\kappa\tan\theta} \, 
\F\bigl(\ts1+\ri\kp-\ri \kpz+i\kappa\tan\theta,\,1+\ri\kp+\ri \kpz+\ri\kappa\tan\theta;
\\ & \qquad\qquad\qquad\qquad\qquad\qquad
\,1+2\ri\kp,\,\threeontwo+\ri\kp-\kappa+\ri\kappa\tan\theta,\,\threeontwo+\ri\kp+\kappa+\ri\kappa\tan\theta;
\,-\zeta^2\sec^2\theta\bigr)\, ,
                                                           \label{u}
\end{split}
\end{equation}
where the $C_i$ are arbitrary constants and the $\F$ are generalized hypergeometric functions \citep{DLMF}. The exact solutions actually date back to \citet{ZhuDzh84aa}, who expressed them in terms of Meijer G-functions. The four solutions become degenerate if $\kappa=0$ or $\kappa_0=0$, where logarithmic solutions are required, but these special cases are easily avoided.

\begin{figure}
\begin{center}
\includegraphics[width=0.35\columnwidth]{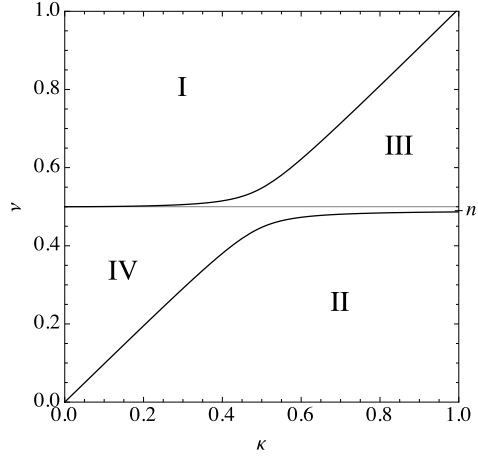}
\caption{Dispersion plane for acoustic gravity waves. The four regions are defined by I: $\kappa_z^2>0$, $\nu>\half$, propagating acoustic waves; II: $\kappa_z^2>0$, $\nu<n$, propagating gravity waves; III: $\kappa_z^2<0$, $\nu>\half$, evanescent acoustic; and IV: $\kappa_z^2<0$, $\nu<\half$, evanescent gravity. Region I lies entirely above the acoustic cutoff frequency $\nu_c=\half$, and Region II is below and asymptotes to the {\BV} frequency $n$. Note that $n\le\half$ with equality only if $\gamma=2$. For $\gamma=\frac{5}{3}$, $n=0.4899$.}
\label{fig:disp}
\end{center}
\end{figure}

Our various dimensionless quantities define separate regions in the $\kappa$-$\nu$ plane, even without the magnetic field. Figure \ref{fig:disp} illustrates these. Region I ($\kappa_z^2>0$, $\nu>\half$) is where acoustic waves may propagate vertically; propagating gravity waves are found in Region II ($\kappa_z^2>0$, $\nu<n$); Regions III and IV ($\kappa_z^2>0$) encompass evanescent waves. The exact solution (\ref{u}) clearly depends on $\kappa_z$, which is the parameter defining these acoustic-gravity regions, but it also depends explicitly on the magnetic field through its inclination $\theta$ and implicitly through $\kappa_0=\sqrt{\nu^2\sec^2\theta-1/4}$. The significance of $\kappa_0$ is that it is real if $\nu\ge\nu_c\cos\theta$, and imaginary otherwise. This represents the ramp effect by which the effective acoustic cutoff frequency is reduced by magnetic field inclination. With inclined magnetic field in place then, Regions II and IV must be split into upper ($\nu>\half\cos\theta$) and lower ($\nu<\half\cos\theta$) sections. The regions are crucial to the correct application of boundary conditions.

The great advantage of the exact solutions is that they prescribe the connectivity between asymptotic behaviours as $z\to\pm\infty$. Conveniently, each of the $\F$ functions approaches 1 as $\zeta\to0$ ($z\to\infty$), and the four constituent terms are respectively the growing exponential fast wave, the decaying (evanescent) fast wave (assuming $\kappa>0$), the upgoing slow wave, and the downgoing slow wave (assuming $\kappa_0$ is real and positive). If $\kappa_0^2<0$ (frequency is below the ramp-reduced acoustic cutoff), the third and fourth terms (apart from the factor $\zeta^{1+2\ri\kappa\tan\theta}$) are decaying and growing exponentials respectively.

Conversely, the $u_i$ can be expressed in terms of an alternate basis \{$U_1$, \ldots, $U_4$\} via $u_j=a_{ij}U_i$, where the summation convention is invoked. Each of the $U_i$ has a pure asymptotic behaviour as $\zeta\to\infty$: $U_1\sim \zeta^{-1/2+2\ri\kappa\tan\theta}e^{2\ri\zeta\sec\theta}$, $U_2\sim\zeta^{-1/2+2\ri\kappa\tan\theta}e^{-2\ri\zeta\sec\theta}$, $U_3\sim\zeta^{-1+2\ri\kappa_z}$, and $U_4\sim\zeta^{-1-2\ri\kappa_z}$ as $\zeta\to\infty$, representing respectively the downgoing slow wave, the upgoing slow wave, the downgoing fast wave, and the upgoing fast wave (assuming $\kappa_z^2>0$). Where $\kappa_z^2<0$, the $U_3$ solution is exponentially decaying with depth and the $U_4$ solution is exponentially growing. The coefficients $a_{ij}$ (see Table \ref{tab:a} in the Appendix) may be written exactly in terms of Gamma functions. Letting $A=(a_{ij})$ be the $4\times4$ matrix of $a$-coefficients and $C=(C_1,\ldots,C_4)^T$ be the coefficient vector, we may write $u=c_i U_i$ where $c=AC$ and $c=(c_1,\ldots,c_4)^T$.

The wave energy flux is made up of acoustic and Poynting components,
\begin{equation}\label{F}
\bmath{F} =  \half\re\left[ p_1 \boldv^*-\frac{1}{\mu}(\boldv^*\vcross\B)\vcross\b \right] ,
\end{equation}
where $p_1$ is the gas pressure perturbation, $\B$ is the equilibrium magnetic field, $\b$ is the perturbed field, $\boldv=u\,\eperp+w\,\epar$ is the plasma velocity with $\epar$ and $\eperp$ the unit vectors parallel to and perpendicular to the field, and the star denotes complex conjugation. The factor of $\half$ results from our harmonic representation and averaging over a period, but is irrelevant in the following. By energy conservation, the vertical component $F_z$ is independent of $z$ (which is confirmed numerically). After much algebra using the asymptotic forms of the solutions, 
the vertical flux
may be calculated as either a Hermitian form in $C$, or one in $c$, as explained in \citet{Cal01aa}. 
Specifically,  $F_z=C^H\Phi C$ which is scaled by $p_\mathrm{mag}/c_s$ where $C$ has dimensions of  velocity, the superscript $H$ represents the Hermitian transpose (conjugate transpose) and
\begin{equation}
\Phi = 
  \begin{pmatrix}
  0 & \frac{\ri\kappa}{\nu} & 0 & 0 \\[2pt]
  -\frac{\ri\kappa}{\nu} & 0 & 0 & 0\\[2pt]
  0 & 0 & \phi_{33}\, \mathcal{U}(\kappa_0^2) & \phi_{34}\,\mathcal{U}(-\kappa_0^2) \\[2pt]
  0 & 0 &  \phi_{34}^*\,\mathcal{U}(-\kappa_0^2) & \phi_{44} \,\mathcal{U}(\kappa_0^2)
  \end{pmatrix}  .  \label{Phi}
\end{equation}
The $\phi_{33} $ and $\phi_{44} $ are real and $\mathcal{U}$ is the unit step function. Alternatively, $F_z=c^H\Psi c$, where $\Phi$ and $\Psi$ are related through $\Phi=A^H\Psi A$. Specifically,
\begin{equation}
\Psi = 
  \begin{pmatrix}
  -\frac{\sec\theta}{\nu} & 0 & 0 & 0 \\[2pt]
  0 & \frac{\sec\theta}{\nu}  & 0 & 0\\[2pt]
  0 & 0 & \psi_{33} \,\mathcal{U}(\kappa_z^2) & \psi_{34}\,\mathcal{U}(-\kappa_z^2) \\[2pt]
  0 & 0 &  \psi_{34}^*\,\mathcal{U}(-\kappa_z^2) & \psi_{44} \,\mathcal{U}(\kappa_z^2)
  \end{pmatrix}  .  \label{Psi}
\end{equation}
The $\phi$ and $\psi$ coefficients are set out in the Appendix. The $\Phi$ form is convenient for calculating fluxes at the top, and the $\Psi$ form is used at the bottom. 

Interpreting the off-diagonal terms requires diagonalization of the matrix. For example, consider $\Psi$ in Region III or IV, where $\kappa_z^2<0$. The total flux is
\begin{equation}
\begin{split}
F_z &=-\frac{\sec\theta}{\nu} \left|C_1\right|^2 + \frac{\sec\theta}{\nu} \left|C_2\right|^2 +C_3^*C_4 \phi_{34} + C_3C_4^*\phi_{34}^* \\
&= -\frac{\sec\theta}{\nu} \left|C_1\right|^2 + \frac{\sec\theta}{\nu} \left|C_2\right|^2 
+ \half \left| C_4+C_3\exp(-\ri\arg\phi_{34})\right|^2 |\phi_{34}| - \half \left| C_4-C_3\exp(-\ri\arg\phi_{34})\right|^2 |\phi_{34}|.
\end{split} \label{Fdiag}
\end{equation}
We see from the first line that there is no flux associated with the off-diagonal terms if either $C_3=0$ or $C_4=0$, i.e., if there is a single exponential decaying or growing evanescent mode. However, if both are present, they carry a net flux. This is rearranged in the second line into upgoing and downgoing components by diagonalization (Hermitian matrices are always diagonalizable). In the case where there is only one evanescent mode, it is conventional to take the up and down fluxes to be zero rather than equal. When they are both present though, in cases where we explore incident flux by tunnelling, the diagonalised terms are identified as reflected and incident fluxes.

Comparing fluxes of incident, reflected, transmitted, and converted waves gives us the various reflection, transmission, and conversion coefficients that are our prime concern here. Details depend upon the specific boundary conditions of each scenario. 

Comparison with the transmission coefficient of Generalized Ray Theory \citep{Cal06aa,SchCal06aa} is enlightening. In its simplest form, the acoustic-to-acoustic transmission coefficient is given by
\begin{equation}
T = \exp\left[-\upi k h_s \sin^2\alpha \right]_{a=c_s}   \label{T}
\end{equation}
evaluated at the Alfv\'en acoustic equipartition level $a=c_s$ (i.e., $\zeta=\nu$). Here $k=|\bmath{k}|$ is the wavenumber, $\alpha$ is the attack angle between the wave vector $\bmath{k}$ and the magnetic field $\B$, and $h_s=[d(a^2/c_s^2)/ds]_{a=c_s}^{-1}$ is the conversion zone thickness as measured along the direction $\hat{\bmath{k}}$. There are more exact (and more complicated) forms, but this will suffice for our purposes. Equation (\ref{T}) was found by \citet{HanCal09aa} to be in good agreement with the exact solutions for small to moderate $\alpha$. The main point here is that the concept of `attack angle' represents a useful physical insight.

We shall now be concerned with two cases: slow waves incident from above, and fast waves incident from above. {In both cases we will be guided by connection diagrams (Figures \ref{Connect2} and \ref{Connect3} to be introduced in Sections \ref{sec:slow} and \ref{sec:fast} respectively). These schematically illustrate the connection between the elementary low-$\beta$ (top) and high-$\beta$ (bottom) asymptotics. For example, in Figure \ref{Connect2} there is an injected (downgoing) slow wave at the top carrying unit flux, a reflected slow wave with flux $\mathscr{R}$,  a transmitted fast wave at the bottom (high-$\beta$) carrying flux $\mathscr{T}$, and a converted slow wave (also at the bottom) carrying flux $\mathscr{C}$. Of course, $\mathscr{R+T+C}=1$ by conservation of energy. There is also an evanescent fast wave at the top that decays exponentially with height and carries no energy in the steady state. The three modes with red crosses (growing fast wave at the top, upgoing fast and slow waves at the bottom) are disallowed by our boundary conditions. The single vertical black line joining the low and high $\beta$ asymptotic regions represents the deep mathematical connections between these asymptotics that are implicit in the hypergeometric solutions, and represented by the connection coefficients $a_{ij}$. Connection diagrams such as these are very useful in sorting out which of the $C_i$ and $c_i$ coefficients are involved in each region of the dispersion diagram.}

\section{Incident Slow Wave from Above} \label{sec:slow}

Each Region, I--IV, including upper and lower parts of II and IV, must be considered separately in applying boundary conditions and writing down the exact solution. 

\begin{figure}    
\begin{center}
\vspace{1cm}
\includegraphics[width=.65\hsize]{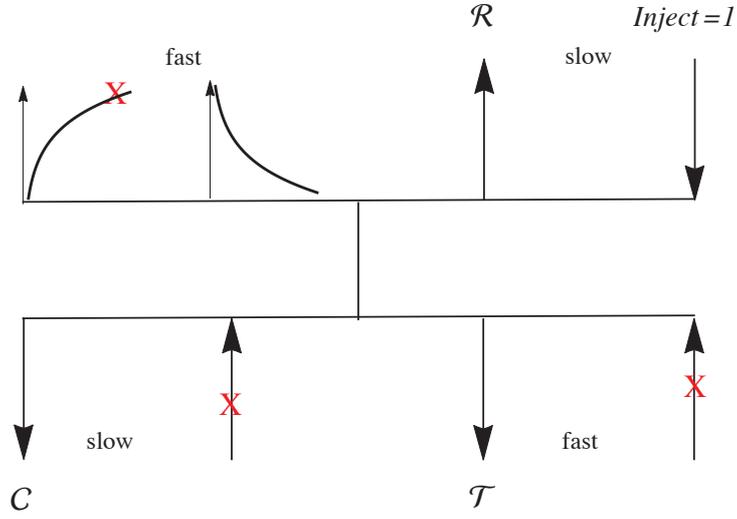}           
\vspace{-1cm}   
\caption{Region I connection diagram for a slow wave incident on the equipartition from above (inject=1).  
{The upper row of arrows represent the low-$\beta$ region waves (from left to right): } the exponentially growing fast wave, the decaying (evanescent) fast wave, the upgoing slow wave and the downgoing slow wave.  {The arrows in the lower row represent  waves in the high-$\beta$ regime (from left to right):} the downgoing slow wave, the upgoing slow wave, the downgoing fast wave and the upgoing fast wave.  The red crosses indicate the elimination of a term due to the application of boundary conditions. ${\cal{T}}$, ${\cal{R}}$ and ${\cal{C}}$ correspond to the transmitted wave, reflected wave and converted wave respectively.}
\label{Connect2}
 \end{center}
 \end{figure}

\begin{description}
\item[{\bf{Region I}}] We set $C_1=0$ (no exponentially growing fast evanescent term at the top), $c_2=0$ (no incoming slow wave at the bottom), and $c_4=0$ (no incoming fast wave at the bottom). With $C_4$ specifying an arbitrary normalisation, $c=AC$ can be solved for the remaining coefficients $C_2$, $C_3$, $c_1$, and $c_3$ (see Figure \ref{Connect2}). The incident (unsigned) flux is then $F_\mathrm{inc}=|C_4|^2|\phi_{44}|$, the reflected flux is $F_\mathrm{ref}=|C_3|^2 |\phi_{33}|$, the transmitted flux (acoustic) is $F_\mathrm{trans}=|c_3|^2 |\psi_{33}|$, and the converted flux (slow magnetic) is $F_\mathrm{conv}=\nu^{-1}|c_1|^2 \sec\theta$. We confirm that $F_\mathrm{ref}+F_\mathrm{trans}+F_\mathrm{conv}=F_\mathrm{inc}$. The reflection, transmission, and conversion coefficients are then defined by
\begin{equation}
\left\{\mathscr{R},\,\mathscr{T},\,\mathscr{C}\right\} = \frac{\left\{F_\mathrm{ref},\,F_\mathrm{trans},\,F_\mathrm{conv}\right\}}{F_\mathrm{inc}},
\end{equation}
with $\mathscr{R}+\mathscr{T}+\mathscr{C}=1$.
\item[{\bf{Region II upper}}] We again set $C_1=0$ and $c_2=0$, but now let $c_3=0$ instead of $c_4$. This is because the flux is reversed in the gravity wave regime (see Equation (\ref{phi33})). All flux formulae are the same as in Region I, except for $F_\mathrm{trans}=|c_4|^2 |\psi_{44}|$.
\item[{\bf{Region II lower}}] Below the ramp frequency $\nu=\half\cos\theta$, where $\kappa_0^2<0$, the bottom right $2\times2$ sub-matrix of $\Phi$ is off-diagonal Hermitian. This is because the `incident' slow wave is now evanescent. Nevertheless, energy can still tunnel through the low-$\beta$ region provided both $C_3$ and $C_4$ are non-zero. We can identify incoming and outgoing components of the tunnelling flux by diagonalization. Specifically, we have
$F_\mathrm{inc}=\half \left| C_4-C_3\exp(-\ri\arg\phi_{34})\right|^2 |\phi_{34}|$, $F_\mathrm{ref}=\half \left| C_4+C_3\exp(-\ri\arg\phi_{34})\right|^2 |\phi_{34}|$, $F_\mathrm{trans}=|c_4|^2 |\psi_{44}|$, and $F_\mathrm{conv}=\nu^{-1}|c_1|^2 \sec\theta$.
\item[{\bf{Region III}}] We set $C_1=c_2=c_4=0$. This time it is the bottom right $2\times2$ sub-matrix of $\Psi$ that is off-diagonal Hermitian, because $\kappa_z$ is imaginary. We have $F_\mathrm{inc}=|C_4|^2|\phi_{44}|$, $F_\mathrm{ref}=|C_3|^2 |\phi_{33}|$, $F_\mathrm{trans}=0$ (since there is no travelling fast mode at the bottom), and $F_\mathrm{conv}=\nu^{-1}|c_1|^2 \sec\theta$.
\item[{\bf{Region IV upper}}] Exactly as for Region III. 
\item[{\bf{Region IV lower}}] Here the bottom right sub-matrices of both $\Phi$ and $\Psi$ are off-diagonal, since both the slow mode at the top and the fast wave at the bottom are evanescent. $C_1=c_2=c_4=0$. We must diagonalize again, to find that
$F_\mathrm{inc}=\half \left|C_4-C_3\exp(-\ri\arg\phi_{34})\right|^2 |\phi_{34}|$, $F_\mathrm{ref}=\half \left|C_4+C_3\exp(-\ri\arg\phi_{34})\right|^2 |\phi_{34}|$, $F_\mathrm{trans}=0$, and $F_\mathrm{conv}=\nu^{-1}|c_1|^2 \sec\theta$.
\end{description}

\begin{landscape}
\begin{figure}
\begin{center}
\includegraphics[width=1.1\textheight]{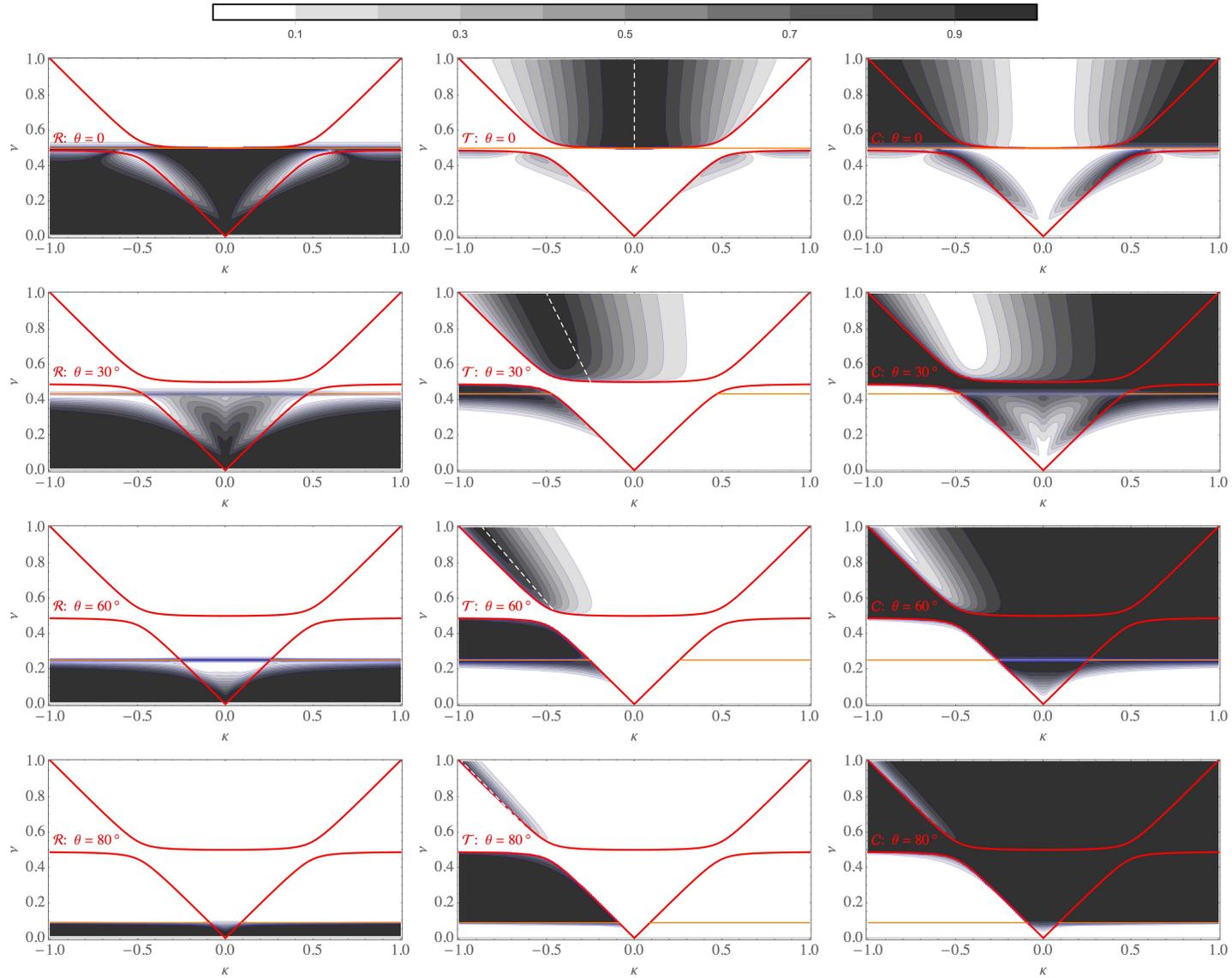}
\caption{Reflection ($\mathscr{R}$, left column), transmission ($\mathscr{T}$, centre column), and conversion ($\mathscr{C}$, right column) coefficients for the case of a slow wave incident from above in magnetic field inclined at angle $0^\circ$ (top), $30^\circ$, $60^\circ$, and $80^\circ$ (bottom) from the vertical. Black corresponds to a coefficient above 0.9, and white to less than 0.1 (see the legend bar at the top). The dashed white line in the Transmission column is $\nu=-\kappa \csc\theta$ and represents the optimal attack angle predicted by Generalized Ray Theory. The transmitted ($\mathscr{T}$) waves are acoustic in nature and can potentially excite seismic events at the photosphere. The heavy red curves separate Regions I, II, III, and IV of acoustic gravity wave theory, and are clearly still relevant in the magnetic case. The horizontal orange line depicts the ramp frequency $\frac{1}{2}\cos\theta$.}
\label{fig:slow}
\end{center}
\end{figure}
\end{landscape}

With these considerations in place, we may calculate $\left\{\mathscr{R},\,\mathscr{T},\,\mathscr{C}\right\}$ for any point in the $\kappa$-$\nu$ plane. Figure \ref{fig:slow} shows the results for $|\kappa|<1$ ($|k_x| <H^{-1}$) and $\nu<1$ ($\omega<2\omega_c$), both very reasonable restrictions for solar waves of seismic interest. In practice, we assume $\kappa>0$ and $-90^\circ<\theta<90^\circ$ in the calculations. However, it is more convenient to present the results graphically with $\kappa$ ranging over both positive and negative values. This is easily done by simply reversing $\theta$ while keeping $\kappa>0$, corresponding to a simple reflection of the $x$-axis.

\begin{figure}
\begin{center}
\includegraphics[width=.7\textwidth]{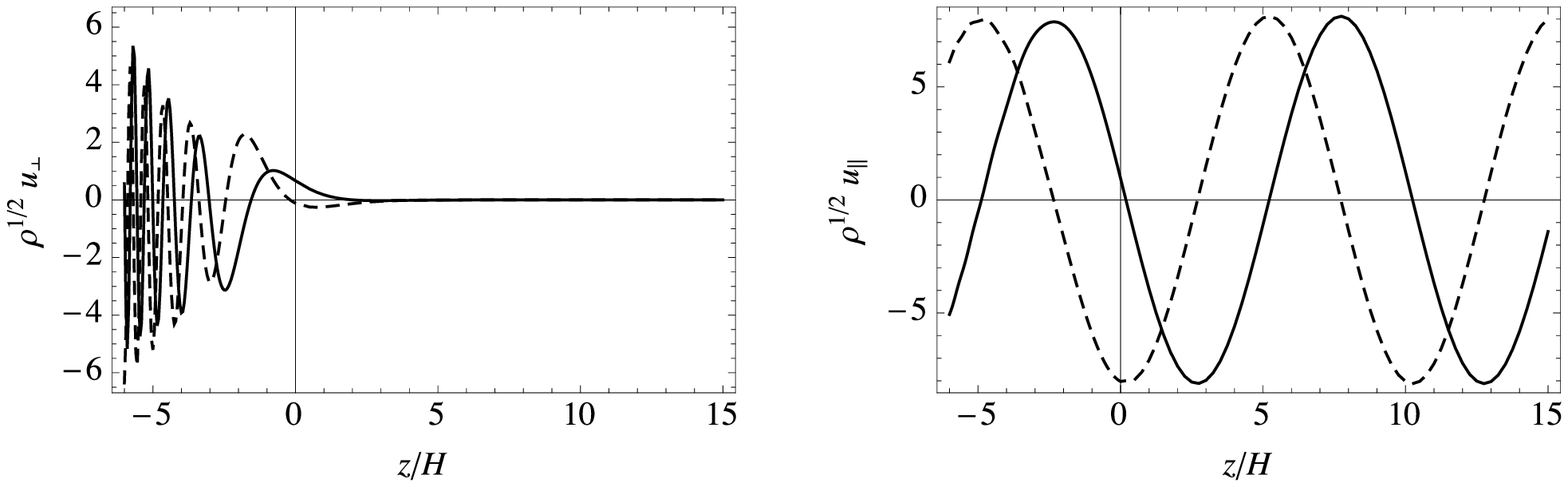}
\includegraphics[width=.7\textwidth]{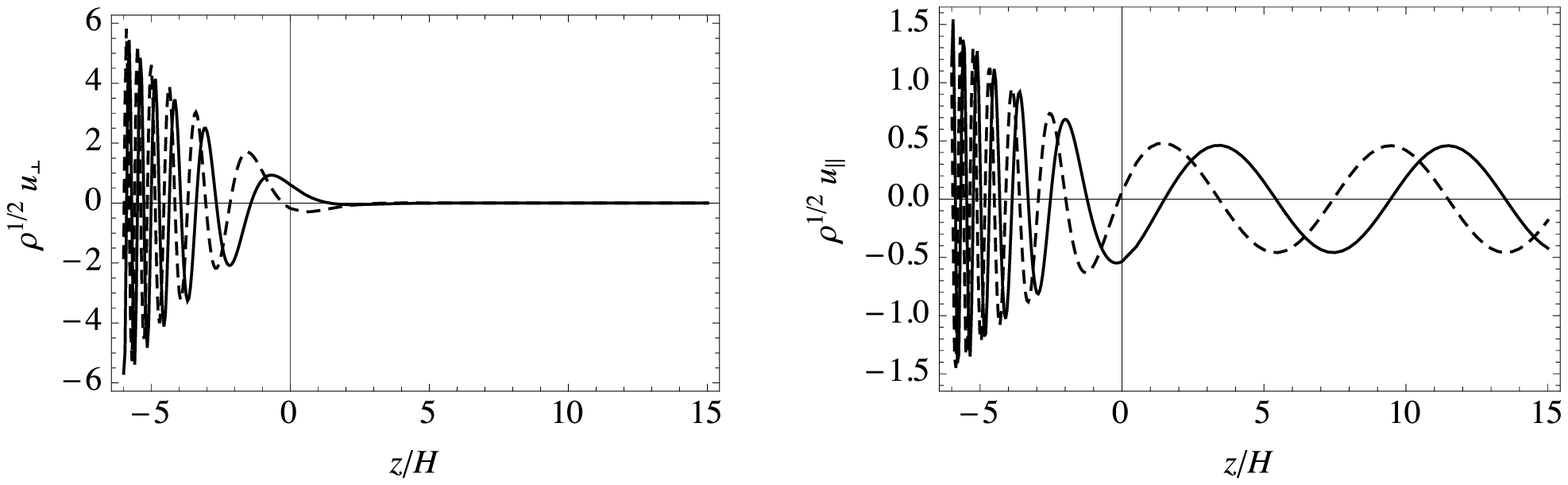}
\caption{Transverse (left column) and longitudinal (right column) velocities as functions of height $z$ for Region I slow waves incident from above. Both are multiplied by the square root of the density to make acoustic wave amplitude uniform for graphical purposes. The real and imaginary parts are rendered in full and dashed curves respectively. Upper row: large transmission case $\kappa=0.55$, $\nu=0.7$, $\theta=30^\circ$, corresponding to $\mathscr{T}=0.85$; Lower row: small transmission case $\kappa=0.6$, $\nu=0.8$, $\theta=30^\circ$, corresponding to $\mathscr{T}=0.01$. Reflection is negligible in both cases. The $-90^\circ$ phase difference between the real and imaginary parts of the slow wave in $z>0$ is characteristic of downgoing waves. The $c_s=a$ equipartition level is placed at $z=0$.}
\label{fig:pltu_slowI}
\end{center}
\end{figure}

Several points warrant our attention.
\begin{enumerate}
\item Black regions in the $\mathscr{T}$ (i.e., second) column represent `gravito-acoustic windows' to the photosphere, where slow waves from above efficiently penetrate $a=c_s$ as fast waves. Similarly, the extensive black regions in the third column are `magnetic windows'.
\item Transmission in Region I is largely restricted to a band about $\nu=-\kappa \csc\theta$ (see the dashed line in Column 2). This is entirely as expected from Generalized Ray Theory (see Equation (\ref{T})), as it corresponds to small attack angle. (Consider horizontal wave number $k_x$ and total wave number $k=\omega/c$, angled downward at angle $\delta$ from the vertical. Then $c k_x/\omega = \kappa/\nu=\sin\delta$. This is along the magnetic field direction if $\delta=-\theta$, i.e., if $\nu=-\kappa\csc\theta$.)
\item There is very little slow wave reflection in Region I (except right on the boundary for $\theta=0^\circ$ and $\kappa\lesssim0.5$). This is consistent with ray theory, where reflection is identically zero. 
\item Plotting the velocities as functions of $z$ for low and high transmission cases (Figure \ref{fig:pltu_slowI} upper and lower rows respectively) makes the nature of transmission very clear. In the high $\mathscr{T}$ case, the longitudinal `acoustic' wave passes through $z=0$ almost unchanged, whereas in the low $\mathscr{T}$ case, it is strongly converted to small wavelength slow (magnetic) waves.
\item There is substantial transmission in the upper part of Region II for $\kappa<0$ and $\theta>0$ (or equivalently $\kappa>0$ and $\theta<0$). That is, travelling incident slow waves with horizontal phase speed less than the sound speed ($|\nu/\kappa|<1$) almost totally transmit to travelling gravity waves at the bottom if directed against the field inclination ($\kappa\sin\theta<0$). There is essentially no transmission if they are directed with the field inclination.
\item Reflection is near-total in Region II below the ramp frequency.
\item Conversion (to slow magnetic waves in $\beta\gg1$) in Region I is almost exactly the complement of transmission. Slow waves, which are asymptotically incompressive as $z\to-\infty$, are unlikely to have the same seismic impact on the photosphere.
\item Conversion also dominates in the upper part of Region II $\kappa>0$.
\item  The top row of Figure \ref{fig:slow} (vertical field) corresponds to the cases considered by \citet{McDHoo07aa}. Although we are unable to compare our results with theirs quantitatively, the qualitative behaviour agrees well for both small and large $\kappa$. Their Figure 3 also shows good agreement with the approximate transmission coefficient \citep[Eqn (26) of][]{SchCal06aa} that we have already shown accords well with the exact solutions \citep{HanCal09aa}.
\end{enumerate}

Overall, incident slow (acoustic) waves at the top are expected to transmit through to the photosphere as acoustic waves only in a narrow band of propagation about the magnetic field direction. Below the {\BV} frequency but above the ramp frequency though, we anticipate almost total transmission as gravity waves. This is most effective at large field inclination as it happens at all frequencies $\half\cos\theta<\nu<n$.

\section{Incident Fast Wave from Above}  \label{sec:fast}
Fast waves at the top are evanescent in all Regions, and so may only transport energy to the $a\approx c$ conversion layer by tunnelling. This time we are most concerned with (magnetic to acoustic) conversion rather than transmission, as this is likely to be most effective seismically.

\begin{figure}    
\begin{center}
\vspace{1cm}
\includegraphics[width=.65\hsize]{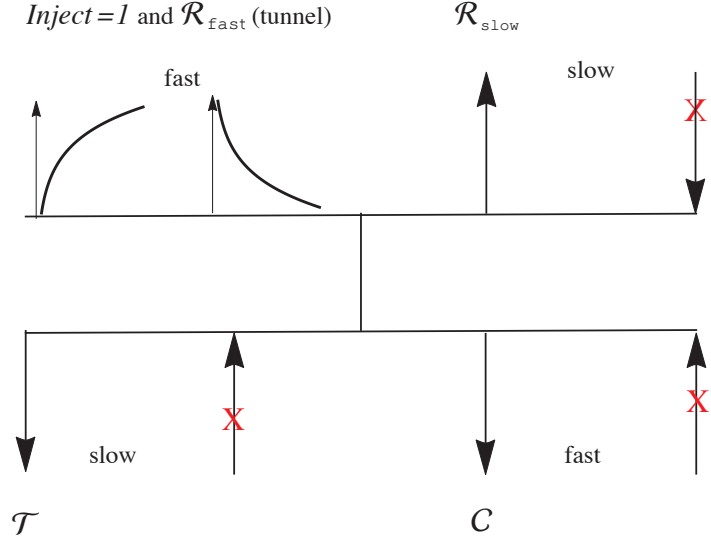}              
\vspace{-1cm}
\caption{Region I connection diagram for a fast wave incident on the equipartition from above (inject=1). {The upper row of arrows represent the low-$\beta$ region waves (from left to right): } the exponentially growing fast wave, the decaying (evanescent) fast wave, the upgoing slow wave and the downgoing slow wave.  {The arrows in the lower row represent  waves in the high-$\beta$ regime (from left to right):} the downgoing slow wave, the upgoing slow wave, the downgoing fast wave and the upgoing fast wave. The red crosses indicate the elimination of a term due to the application of boundary conditions. ${\cal{T}}$, ${\cal{R}}_{\rm{fast}}$, ${\cal{R}}_{\rm{slow}}$ and ${\cal{C}}$ correspond to the transmitted wave, reflected fast wave, reflected slow wave and the converted wave respectively.}
\label{Connect3}
 \end{center}
 \end{figure}
 
\begin{description}
\item[{\bf{Region I}}] We set $C_4=0$ (no incoming slow wave at the top), $c_2=0$ (no incoming slow wave at the bottom), and $c_4=0$ (no incoming fast wave at the bottom).  With $C_1$ specifying an arbitrary normalisation we once again solve $c=AC$ for the remaining coefficients $C_1, C_2, C_3, c_1$ and $c_3$ (see Figure \ref{Connect3}).  
We can identify the incoming and outgoing components of the tunnelling flux by diagonalization.  The incident and reflected fast fluxes are $F_{\rm{inc}}=\frac{1}{2}|K_1|^2|\phi_{12}|$ and $F_{\rm{ref,f}}=\frac{1}{2}|K_2|^2|\phi_{12}|$ respectively.   In this case we have two types of reflection, the fast wave $F_{\rm{ref,f}}$ associated with tunnelling and the normal slow mode $F_{\rm{ref,s}}=|C_3|^2\phi_{33}$. 
Transmission is to a downward slow (magnetic) wave $F_{\rm{trans}}=|c_1|^2 \psi_{11}$ and  conversion to a downgoing fast (acoustic) wave $F_{\rm{conv}}=|c_3|^2 \psi_{33}$.  
\item[{\bf{Region II upper}}] Same as Region I with $C_4=c_2=0$ except now $c_3=0$ instead of $c_4$ as the  converted fast (gravity) wave, $F_{\rm{conv}}=|c_4|^2 \psi_{44}$ will be oppositely directed.
 \item[{\bf{Region II lower}}] Same as Region II upper except with $c_3=0$ as there is no reflected slow (acoustic) wave $F_{\rm{ref,s}}=0$.
\item[{\bf{Region III}}] Same as Region I except with $c_4=0$ as there is no fast (acoustic) converted wave, $F_{\rm{conv}}=0$.
\item[{\bf{Region IV upper}}] Same as Region III except $F_{\rm{trans}}=0$ as there is no slow (magnetic) transmitted flux. 
\item[{\bf{Region IV lower}}]  Here there is no fast (acoustic) converted wave \emph{i.e.,} $F_{\rm{conv}}=0$ and no slow (acoustic) reflected wave ($F_{\rm{ref,s}}=0$) so that, $C_3=C_4=c_2=c_3=c_4=0$.  
\end{description}

\begin{landscape}
\begin{figure}
\begin{center}
\includegraphics[width=1.3\textheight]{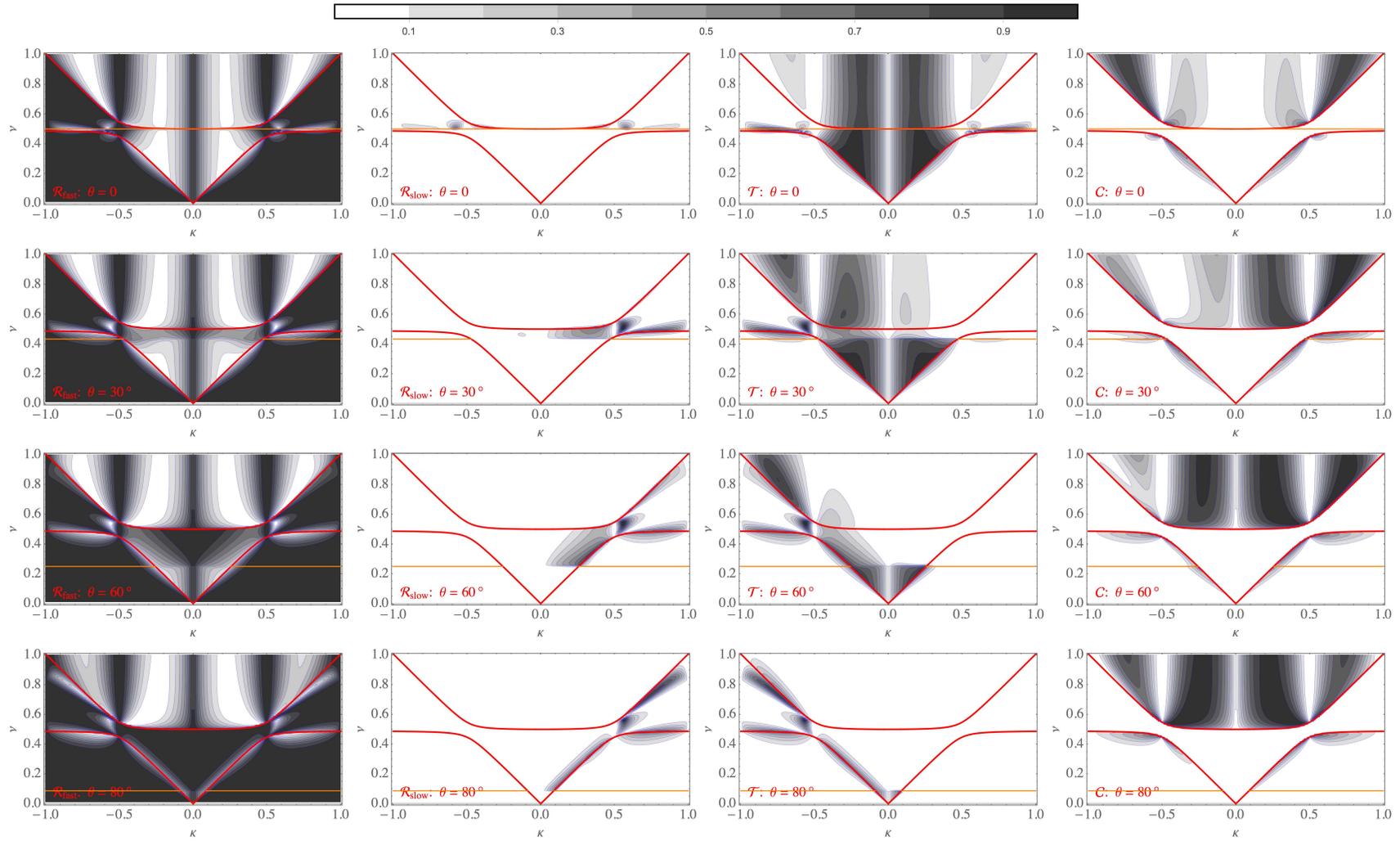}
\caption{Fast wave reflection ($\mathscr{R}_\mathrm{\,fast}$, left column), slow wave reflection ($\mathscr{R}_\mathrm{\,slow}$, second column), transmission ($\mathscr{T}$, third column), and conversion ($\mathscr{C}$, right column) coefficients for the case of an evanescent fast wave incident from above in magnetic field inclined at angle $0^\circ$ (top), $30^\circ$, $60^\circ$, and $80^\circ$ (bottom) from the vertical. The converted ($\mathscr{C}$) waves are acoustic in nature and can potentially excite seismic events at the photosphere.}
\label{fig:fast}
\end{center}
\end{figure}
\end{landscape}

Subject to these restrictions the coefficients  $\left\{\mathscr{R}_{\,\rm{fast}},\, \mathscr{R}_{\rm{\,slow}},\,\mathscr{T},\,\mathscr{C}\right\}$ are calculated with the results shown in Figure \ref{fig:fast} for the $\kappa$-$\nu$ plane with  $|\kappa|<1$ ($|k_x| <H^{-1}$) and $\nu<1$ ($\omega<2\omega_c$).  The main points of interest are briefly outlined below.

 \begin{enumerate}
  \item Fast wave reflection appears to be symmetrical about $\kappa=0$ (vertical wave).  Reflection is  almost total in Regions IIa, IIb and III and increases in Regions I, IVa and IVb  as the magnetic field inclination increases. 
  \item   Minimal slow wave reflection, mostly about $\kappa_z=0$ and in Region IV.   All wave energy flux is above the (ramp reduced) acoustic cut-off frequency and there is none for Regions I, IIa and IIb.
 \item The amount of transmission to the slow (magnetic) wave decreases as the magnetic field inclination increases. 
 \item  In Region I where acoustic waves are propagating we see that a substantial fraction of the incident fast wave converts to an acoustic wave in the high-$\beta$ regime.
\item There is no conversion in Regions III and IV.  
 \end{enumerate}
 
\section{Conclusions}
     Seismic transients to date are only detected within sunspot penumbra, the magnetic field of which is highly inclined.  Since transmission is selective, with a dependence on alignment between wavevector and field inclination, the coupling or tunnelling of waves through the $a=c_s$ equipartition may explain the compact nature of sunquakes.   Current simulations do not model regions of strong, oblique fields \citep{LinDonMar14aa} even though it may prove to be an important factor. Of course, flare waves will not be linear in most cases, and our model is too simplistic in many other ways. However, the study at least presents some concepts that may play a role in the final theory of sunquake excitation.
 
     Our results indicate that for frequencies above the acoustic cutoff and in highly inclined magnetic fields, the slow waves generated in the low-$\beta$ regime are more susceptible to conversion.  Is it possible for these transverse field-guided magnetic waves in the high-$\beta$ regime to be sufficiently focussed to constitute the quake source?   \cite{JudKleDon14aa} indeed found a weak but significant seismic source between two leading sunspots, closer to a magnetic pore, where the magnetic field configuration could be significantly  horizontal.  In any case, it should be noted that the incident fast wave also tends to favour conversion for large field inclinations, which excites fast (acoustic) waves below the $a=c_s$ equipartition.
 
     Every solar flare seems to be different. Perhaps the low chromosphere acts as a moderator allowing a certain mechanism to dominate, dependent on local conditions. For example, the dominance of the transmission and/or reflection and/or conversion coefficients could respond to the local physics in ways that our model may hint at. High resolution data in the future will allow a more detailed analysis of acoustic-wave propagation and dissipation effects in the solar chromosphere. Despite the simplicity of the uniform field isothermal model, which of course is unable to explain the complexities of sunquake excitation, it may be helpful in identifying some physics which is relevant to the process.
     
{     The true source of sunquakes appears to be a short lived impulse, rather than a mono-frequency wave as examined here. We should therefore expect a broad distribution of frequencies in the driver, and probably a variable frequency response from the photosphere. \cite{AlvBuiMar12aa} (Figure 1) find a seismic emission energy spectrum from the flare of 15 February 2011 in NOAA AR11158 that falls exponentially with frequency from 3 mHz to 8 mHz. Perhaps this sunquake energy profile is inherent in the flare source -- that is unknown at present -- or perhaps it is a consequence of a filtering effect of the atmosphere. Answering that question definitively is beyond the capabilities of the authors at present. However, the very different transmission and conversion behaviours for slow (acoustic) waves incident from above depending on whether the frequency is above or below the acoustic cutoff of the atmosphere ($\nu=\half$ in dimensionless units; about 5 mHz typically) is very suggestive. Referring to Fig.~\ref{fig:slow}, middle column, it is clear that above the cutoff, only slow waves with wavevectors nearly parallel to the magnetic field can penetrate (transmit) to the photosphere as acoustic waves, especially at large field inclination $\theta$. On the other hand, below $\omega_c$ but above $\omega_c\cos\theta$ (i.e., $\half\cos\theta<\nu<\half$), transmission is almost total for $\kappa<0$. In that sense, canopy magnetic field acts as an imperfect low-pass filter, and may help explain the spectrum reported by \cite{AlvBuiMar12aa}.  At frequencies above $\nu=\half\cos\theta$, there is little reflection at any field inclination, so the bulk of the complementary (converted) flux takes the form of the (magnetic) slow wave below $a=c_s$, with presumably reduced seismic repercussions. 
     }



\bibliographystyle{mnras}        
\bibliography{fred
}

\begin{table}
\caption{Coefficients $a_{ij}$ connecting the $u_j$ and $U_i$ basis functions via $u_j=a_{ij}U_i$ (with summation convention).\label{tab:a}}
\begin{tabular}{ccc}
\hline
$i$ & $j$ & $a_{ij}$\\
\hline
1 & 1 & $\displaystyle \frac{e^{\frac{3 \ri \upi }{4}-\upi  \kappa  (\tan \theta\,-\ri)} \kappa\,\Gamma (-2 \kappa ) \Gamma \left(-\ri \tan \theta\, \kappa
   -\kappa -\ri \kappa _0+\frac{1}{2}\right) \Gamma \left(-\ri \tan \theta\, \kappa -\kappa +\ri \kappa _0+\frac{1}{2}\right) \sec ^{2 \ri \tan
   \theta\, \kappa +2 \kappa -\frac{1}{2}}\theta\,}{\sqrt{\upi }\, \Gamma \left(-\kappa -\ri \kappa _z+\frac{1}{2}\right) \Gamma \left(-\kappa +\ri
   \kappa _z+\frac{1}{2}\right)} $ \\[10pt]
 1 & 2 & $\displaystyle -\frac{e^{\frac{3 \ri \upi }{4}-\upi  \kappa  (\tan \theta\,+\ri)} \kappa\,\Gamma (2 \kappa ) \Gamma \left(-\ri \tan \theta\, \kappa
   +\kappa -\ri \kappa _0+\frac{1}{2}\right) \Gamma \left(-\ri \tan \theta\, \kappa +\kappa +\ri \kappa _0+\frac{1}{2}\right) \sec ^{2 \ri \tan
   \theta\, \kappa -2 \kappa -\frac{1}{2}}\theta\,}{\sqrt{\upi }\, \Gamma \left(\kappa -\ri \kappa _z+\frac{1}{2}\right) \Gamma \left(\kappa +\ri
   \kappa _z+\frac{1}{2}\right)} $ \\[10pt]
 1 & 3 & $\displaystyle -\frac{e^{\frac{\ri \upi }{4}-\upi  \kappa _0} \Gamma \left(1-2 \ri \kappa _0\right) \Gamma \left(\ri \tan \theta\, \kappa -\kappa -\ri \kappa
   _0+\frac{3}{2}\right) \Gamma \left(\ri \tan \theta\, \kappa +\kappa -\ri \kappa _0+\frac{3}{2}\right) \sec ^{2 \ri \kappa _0-\frac{3}{2}}\theta\,
   }{2 \sqrt{\upi }\, \Gamma \left(-\ri \left(\kappa _0+\kappa _z-\kappa  \tan \theta\,+\ri\right)\right) \Gamma \left(-\ri \kappa _0+\ri \kappa _z+\ri
   \kappa  \tan \theta\,+1\right)} $ \\[10pt]
 1 & 4 & $\displaystyle -\frac{e^{\upi  \kappa _0+\frac{\ri \upi }{4}} \Gamma \left(2 \ri \kappa _0+1\right) \Gamma \left(\ri \tan \theta\, \kappa -\kappa +\ri \kappa
   _0+\frac{3}{2}\right) \Gamma \left(\ri \tan \theta\, \kappa +\kappa +\ri \kappa _0+\frac{3}{2}\right) \sec ^{-2 \ri \kappa
   _0-\frac{3}{2}}\theta\,}{2 \sqrt{\upi }\, \Gamma \left(\ri \left(\kappa _0-\kappa _z+\kappa  \tan \theta\,-\ri\right)\right) \Gamma \left(\ri
   \left(\kappa _0+\kappa _z+\kappa  \tan \theta\,-\ri\right)\right)} $ \\[10pt] 
 2 & 1 & $\displaystyle -\frac{e^{\upi  \kappa  (\tan \theta\,-\ri)+\frac{\ri \upi }{4}} \kappa\,\Gamma (-2 \kappa ) \Gamma \left(-\ri \tan \theta\, \kappa -\kappa
   -\ri \kappa _0+\frac{1}{2}\right) \Gamma \left(-\ri \tan \theta\, \kappa -\kappa +\ri \kappa _0+\frac{1}{2}\right) \sec ^{2 \ri \tan \theta\,
   \kappa +2 \kappa -\frac{1}{2}}\theta\,}{\sqrt{\upi }\, \Gamma \left(-\kappa -\ri \kappa _z+\frac{1}{2}\right) \Gamma \left(-\kappa +\ri \kappa
   _z+\frac{1}{2}\right)} $ \\[10pt]
 2 & 2 & $\displaystyle \frac{e^{\upi  \kappa  (\tan \theta\,+\ri)+\frac{\ri \upi }{4}} \kappa\,\Gamma (2 \kappa ) \Gamma \left(-\ri \tan \theta\, \kappa +\kappa
   -\ri \kappa _0+\frac{1}{2}\right) \Gamma \left(-\ri \tan \theta\, \kappa +\kappa +\ri \kappa _0+\frac{1}{2}\right) \sec ^{2 \ri \tan \theta\,
   \kappa -2 \kappa -\frac{1}{2}}\theta\,}{\sqrt{\upi }\, \Gamma \left(\kappa -\ri \kappa _z+\frac{1}{2}\right) \Gamma \left(\kappa +\ri \kappa
   _z+\frac{1}{2}\right)} $ \\[10pt]
 2 & 3 & $\displaystyle \frac{e^{\upi  \kappa _0+\frac{3 \ri \upi }{4}} \Gamma \left(1-2 \ri \kappa _0\right) \Gamma \left(\ri \tan \theta\, \kappa -\kappa -\ri
   \kappa _0+\frac{3}{2}\right) \Gamma \left(\ri \tan \theta\, \kappa +\kappa -\ri \kappa _0+\frac{3}{2}\right) \sec ^{2 \ri \kappa
   _0-\frac{3}{2}}\theta\,}{2 \sqrt{\upi }\, \Gamma \left(-\ri \left(\kappa _0+\kappa _z-\kappa  \tan \theta\,+\ri\right)\right) \Gamma \left(-\ri
   \kappa _0+\ri \kappa _z+\ri \kappa  \tan \theta\,+1\right)} $ \\[10pt]
 2 & 4 & $\displaystyle \frac{e^{\frac{3 \ri \upi }{4}-\upi  \kappa _0} \Gamma \left(2 \ri \kappa _0+1\right) \Gamma \left(\ri \tan \theta\, \kappa -\kappa +\ri
   \kappa _0+\frac{3}{2}\right) \Gamma \left(\ri \tan \theta\, \kappa +\kappa +\ri \kappa _0+\frac{3}{2}\right) \sec ^{-2 \ri \kappa
   _0-\frac{3}{2}}\theta\,}{2 \sqrt{\upi }\, \Gamma \left(\ri \left(\kappa _0-\kappa _z+\kappa  \tan \theta\,-\ri\right)\right) \Gamma \left(\ri
   \left(\kappa _0+\kappa _z+\kappa  \tan \theta\,-\ri\right)\right)} $ \\[10pt]  
 3 & 1 & $\displaystyle \frac{\Gamma (1-2 \kappa ) \Gamma \left(2 \ri \kappa _z\right) \Gamma \left(-\ri \tan \theta\, \kappa -\kappa -\ri \kappa
   _0+\frac{1}{2}\right) \Gamma \left(-\ri \tan \theta\, \kappa -\kappa +\ri \kappa _0+\frac{1}{2}\right) \sec ^{2 \kappa +2 \ri \kappa
   _z-1}\theta\,}{\Gamma \left(-\kappa +\ri \kappa _z+\frac{1}{2}\right)^2 \Gamma \left(i \left(\kappa _0+\kappa _z-\kappa  \tan \theta\,
   \right)\right) \Gamma \left(-\ri \left(\kappa _0-\kappa _z+\kappa  \tan \theta\,\right)\right)} $ \\[10pt]
 3 & 2 & $\displaystyle \frac{\Gamma (2 \kappa +1) \Gamma \left(2 \ri \kappa _z\right) \Gamma \left(-\ri \tan \theta\, \kappa +\kappa -\ri \kappa
   _0+\frac{1}{2}\right) \Gamma \left(-\ri \tan \theta\, \kappa +\kappa +\ri \kappa _0+\frac{1}{2}\right) \sec ^{-2 \kappa +2 \ri \kappa
   _z-1}\theta\,}{\Gamma \left(\kappa +\ri \kappa _z+\frac{1}{2}\right)^2 \Gamma \left(\ri \left(\kappa _0+\kappa _z-\kappa  \tan \theta\,
   \right)\right) \Gamma \left(-\ri \left(\kappa _0-\kappa _z+\kappa  \tan \theta\,\right)\right)} $ \\[10pt]
 3 & 3 & $\displaystyle \frac{\Gamma \left(1-2 \ri \kappa _0\right) \Gamma \left(2 \ri \kappa _z\right) \Gamma \left(\ri \tan \theta\, \kappa -\kappa -\ri \kappa
   _0+\frac{3}{2}\right) \Gamma \left(\ri \tan \theta\, \kappa +\kappa -\ri \kappa _0+\frac{3}{2}\right) \sec ^{2 \ri \kappa _0+2 \ri \kappa _z-2 \ri
   \kappa  \tan \theta\,-2}\theta\,}{\Gamma \left(-\kappa +\ri \kappa _z+\frac{1}{2}\right) \Gamma \left(\kappa +\ri \kappa
   _z+\frac{1}{2}\right) \Gamma \left(-\ri \kappa _0+\ri \kappa _z+\ri \kappa  \tan \theta\,+1\right) \Gamma \left(-\ri \left(\kappa _0-\kappa
   _z+\kappa  \tan \theta\,\right)\right)} $ \\[10pt]
 3 & 4 & $\displaystyle \frac{\Gamma \left(2 \ri \kappa _0+1\right) \Gamma \left(2 \ri \kappa _z\right) \Gamma \left(\ri \tan \theta\, \kappa -\kappa +\ri \kappa
   _0+\frac{3}{2}\right) \Gamma \left(\ri \tan \theta\, \kappa +\kappa +\ri \kappa _0+\frac{3}{2}\right) \sec ^{-2 \ri \kappa _0+2 \ri \kappa _z-2 \ri
   \kappa  \tan \theta\,-2}\theta\,}{\Gamma \left(-\kappa +\ri \kappa _z+\frac{1}{2}\right) \Gamma \left(\kappa +\ri \kappa
   _z+\frac{1}{2}\right) \Gamma \left(\ri \left(\kappa _0+\kappa _z-\kappa  \tan \theta\,\right)\right) \Gamma \left(\ri \left(\kappa _0+\kappa
   _z+\kappa  \tan \theta\,-\ri\right)\right)} $ \\[10pt]  
 4 & 1 & $\displaystyle \frac{\Gamma (1-2 \kappa ) \Gamma \left(-2 \ri \kappa _z\right) \Gamma \left(-\ri \tan \theta\, \kappa -\kappa -\ri \kappa
   _0+\frac{1}{2}\right) \Gamma \left(-\ri \tan \theta\, \kappa -\kappa +\ri \kappa _0+\frac{1}{2}\right) \sec ^{2 \kappa -2 \ri \kappa
   _z-1}\theta\,}{\Gamma \left(-\kappa -\ri \kappa _z+\frac{1}{2}\right)^2 \Gamma \left(-\ri \left(-\kappa _0+\kappa _z+\kappa  \tan \theta\,
   \right)\right) \Gamma \left(-\ri \left(\kappa _0+\kappa _z+\kappa  \tan \theta\,\right)\right)} $ \\[10pt]
 4 & 2 & $\displaystyle \frac{\Gamma (2 \kappa +1) \Gamma \left(-2 \ri \kappa _z\right) \Gamma \left(-\ri \tan \theta\, \kappa +\kappa -\ri \kappa
   _0+\frac{1}{2}\right) \Gamma \left(-\ri \tan \theta\, \kappa +\kappa +\ri \kappa _0+\frac{1}{2}\right) \sec ^{-2 \kappa -2 \ri \kappa
   _z-1}\theta\,}{\Gamma \left(\kappa -\ri \kappa _z+\frac{1}{2}\right)^2 \Gamma \left(-\ri \left(-\kappa _0+\kappa _z+\kappa  \tan \theta\,
   \right)\right) \Gamma \left(-\ri \left(\kappa _0+\kappa _z+\kappa  \tan \theta\,\right)\right)} $ \\[10pt]
 4 & 3 & $\displaystyle \frac{\Gamma \left(1-2 \ri \kappa _0\right) \Gamma \left(-2 \ri \kappa _z\right) \Gamma \left(\ri \tan \theta\, \kappa -\kappa -\ri \kappa
   _0+\frac{3}{2}\right) \Gamma \left(\ri \tan \theta\, \kappa +\kappa -\ri \kappa _0+\frac{3}{2}\right) \sec ^{2 \ri \kappa _0-2 \ri \kappa _z-2 \ri
   \kappa  \tan \theta\,-2}\theta\,}{\Gamma \left(-\kappa -\ri \kappa _z+\frac{1}{2}\right) \Gamma \left(\kappa -\ri \kappa
   _z+\frac{1}{2}\right) \Gamma \left(-\ri \left(\kappa _0+\kappa _z-\kappa  \tan \theta\,+\ri\right)\right) \Gamma \left(-\ri \left(\kappa
   _0+\kappa _z+\kappa  \tan \theta\,\right)\right)} $ \\[10pt]
 4 & 4 & $\displaystyle \frac{\Gamma \left(2 \ri \kappa _0+1\right) \Gamma \left(-2 \ri \kappa _z\right) \Gamma \left(\ri \tan \theta\, \kappa -\kappa +\ri \kappa
   _0+\frac{3}{2}\right) \Gamma \left(\ri \tan \theta\, \kappa +\kappa +\ri \kappa _0+\frac{3}{2}\right) \sec ^{-2 \ri \kappa _0-2 \ri \kappa _z-2 \ri
   \kappa  \tan \theta\,-2}\theta\,}{\Gamma \left(-\kappa -\ri \kappa _z+\frac{1}{2}\right) \Gamma \left(\kappa -\ri \kappa
   _z+\frac{1}{2}\right) \Gamma \left(i \left(\kappa _0-\kappa _z+\kappa  \tan \theta\,-\ri\right)\right) \Gamma \left(-\ri \left(-\kappa
   _0+\kappa _z+\kappa  \tan \theta\,\right)\right)} $ \\
\end{tabular}
\end{table}

\appendix

\section{Practicalities of Calculating with the Exact Solutions}

A \emph{Mathematica} notebook  is provided as supplementary material, containing all necessary utilities to calculate and plot $\mathscr{R}$, $\mathscr{T}$, and $\mathscr{C}$. The $A$, $\Phi$ and $\Psi$ matrices are included, and the waves are also animated. EMAGWIA.nb (Exact Magneto-Acoustic-Gravity Waves in Isothermal Atmospheres) is written in \emph{Mathematica} 10, and contains full instructions for its use. 

\subsection{$A$-coefficients}

The coupling coefficients $a_{ij}$ are set out in Table \ref{tab:a}. These completely specify how the fast and slow modes in the low-$\beta$ limit couple to those in the high-$\beta$ limit. They therefore contain all information about mode conversion and transmission.

\subsection{Flux Matrix Components}
The $\psi_{ij}$ terms in the flux matrix $\Psi$ can be written out fairly compactly:
\begin{equation}
\psi_{33}=
-\frac{\nu  \kappa _z}{\nu ^2-n^2}
 \left| \frac{ n^2 \gamma  \sin \theta  \left(\ri (\gamma -2)+2 \gamma  \kappa
   _z\right) \nu ^2-2 (\gamma -1) \kappa  (\nu^2-n^2) \cos \theta }{2 (\gamma -1) \left(\left(\nu
   ^2-n^2\right) \kappa ^2+\left(n^2 \kappa ^2-\nu ^4\right) \sin ^2\theta \right)-\ri n^2 (\gamma -2) \gamma  \kappa  \nu
   ^2 \sin 2 \theta }\right|^2;  \label{phi33}
\end{equation}
$\psi_{44}$ is exactly the same except with $\kappa_z$ replaced by $-\kappa_z$ throughout. Note that $\sgn\psi_{33}=-\sgn(\nu^2-n^2)$ and $\sgn\psi_{44}=+\sgn(\nu^2-n^2)$, so the `3,3' term relates to the downgoing fast wave in Region I and the upgoing wave in Region II. The `4,4' term is the converse.

The off-diagonal term, which only comes into play when $\kappa_z$ is imaginary, is
\begin{equation}
\psi_{34}=
\frac{\ri \nu  \left| \kappa _z\right|  f}{(\nu^2-n^2)g},
\end{equation}
where
\begin{multline}
f=
4 (\gamma -1)^2 \kappa ^2 \cos ^2\theta  \left(n^2-\nu
   ^2\right)^2\\
   +\nu ^2 \sin \theta  \left[\gamma  \nu ^2 n^2 \sin \theta  \left(2 \gamma  \left| \kappa
   _z\right| +\gamma -2\right) \left((\gamma -2) \gamma  n^2-2 (\gamma -1) \left| \kappa _z\right|
   \right)
   -8 \ri (\gamma -1)^2 \kappa  \cos \theta  (n^2-\nu^2 ) \left| \kappa _z\right|
   \right]
\end{multline}
and
\begin{equation}
g= (\gamma -2)^2 \gamma ^2 \kappa ^2 \nu ^4 n^4 \sin ^2 2 \theta +(\gamma -1)^2
   \left(-2 \kappa ^2 \nu ^2+\nu ^4+ \left(\kappa ^2 n^2-\nu ^4\right)\cos 2 \theta +\kappa ^2 n^2\right)^2 .
\end{equation}

The $\phi$-coefficients have been worked out, but their expressions are too long to present here. They are written out explicitly in the accompanying \emph{Mathematica} notebook. They can also be recovered from $\Psi$ using $\Phi=A^H\Psi A$, though in practice we calculated them independently and only used the relationship with $\Psi$ as a check.

\end{document}